# Performance Expectations for a Tomography System Using Cosmic Ray Muons and Micro Pattern Gas Detectors for the Detection of Nuclear Contraband

Kondo Gnanvo, *Member, IEEE*, Patrick Ford, Jennifer Helsby, Richie Hoch, Debasis Mitra, *Senior Member, IEEE*, Marcus Hohlmann, *Member, IEEE*

*Abstract*–We present results from a detailed GEANT4 simulation of a proposed Muon Tomography System that employs compact Micro Pattern Gas Detectors with high spatial resolution. A basic Point-Of-Closest-Approach algorithm is applied to reconstructed muon tracks for forming 3D tomographic images of interrogated targets. Criteria for discriminating materials by Z and discrimination power achieved by the technique for simple scenarios are discussed for different integration times. The simulation shows that Muon Tomography can clearly distinguish high-Z material from low-Z and medium-Z material. We have studied various systematic effects that affect the performance of the MT and the discrimination power. The implications of the simulation results for the planned development of a prototype MT station are discussed.

## I. INTRODUCTION

Standard radiation detection techniques currently employed by radiation portal monitors at international borders and ports are not very sensitive to high-Z radioactive material, e.g. U or Pu, if the material is well shielded to absorb the emanating radiation. Muon Tomography (MT) based on the measurement of multiple scattering [1] of atmospheric cosmic ray muons traversing cargo or vehicles is a promising technique for solving this problem [2-6] because of the deep penetration of cosmic ray muons into shielding material. The deviation of a muon track due to multiple scattering depends on the Z of the material traversed by the track. The technique exploits this dependence for discriminating materials by Z.

We have performed detailed GEANT4 simulations of a proposed MT system that employs compact Micro Pattern Gas Detectors (MPGD) with high spatial resolution, e.g. GEM detectors [7]. A basic Point-Of-Closest-Approach (POCA) algorithm [8] is applied to reconstructed muon tracks for forming 3D tomographic images of interrogated targets. The geometric acceptance, basic Z-discrimination and imaging capabilities, and performance for shielded targets have been discussed elsewhere [9]. Here we present a statistical analysis of the Z discrimination power achievable with the technique and the impact of systematic effects – in particular spatial resolution of the MPGD.

## II. SIMULATION OF THE MUON TOMOGRAPHY STATION

A typical geometry of a Muon Tomography Station for our simulation is shown in Fig. 1. The MT station comprises tracking stations at the top, the bottom, and on the sides. Each tracking station is made of three GEM detectors (or GEMs) spaced 5 mm apart. The top and bottom tracking stations have an area of 4 m × 4 m with a 3 m gap between top and bottom planes. Inside the MT we place five material blocks, i.e. targets to be interrogated, of different Z values - low-Z Al, medium-Z Fe, and high-Z Pb, W, U, within the horizontal plane at z = 0. The typical target is 40 cm wide, 40 cm long, and 10 cm thick.

For the simulation, we use the Monte Carlo generator software package CRY [10], [11] to generate muons with angular distribution and energy spectrum corresponding to those of cosmic ray muons at sea level, within a 10 × 10 m² horizontal plane above the MT station. The generator is interfaced with the GEANT4 simulation package [12], [13] for simulating the geometry of the MT station, the tracking of muons, and the interaction of the muons with targets inside the MT station including multiple scattering. A run of 10 million GEANT4 events is equivalent to about 10 min exposure of the MT to cosmic ray muons under real conditions.

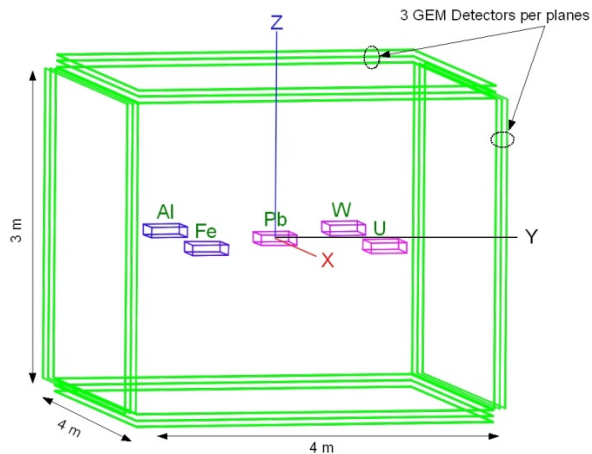

Fig. 1: The geometry of a basic MT scenario with five targets from low-Z material (Al) to high-Z (U) inside. The target dimensions are 40 × 40 × 10 cm³.

## A. Muon Tracking

Every muon traversing a pair of tracking station (top/bottom), (top/lateral), (lateral/bottom) is accepted for probing the MT volume. Incoming and outgoing muon tracks are obtained by least-squares fitting of the hits recorded by the 3 GEMs of the corresponding tracking station. The spatial resolution of the GEM detectors is simulated by smearing simultaneously the x and y coordinates of these hits using Gaussians with widths ($\sigma$) of 50 µm, 100 µm, and 200 µm, respectively.

## B. Scattering angle distributions and statistics

The scattering angle is calculated from the incoming and outgoing GEANT4 muon tracks. We study the scattering angle distributions for targets with different Z-values and the effect of the detector resolution on these distributions. We generate large data samples of more than 1 million muon tracks for each distribution to achieve good statistical precision.

Fig. 2 shows that the effect of the detector resolution is to shift the distribution towards higher value. The mean scattering angle as well as the width of the distribution (rms) increases with the resolution.

In Fig. 3, we plot the mean scattering angle and the rms against the spatial resolution. The plots at the top show results for the standard MT geometry with a 5 mm gap between the GEMs for each tracking station. The difference between the mean scattering angle (and rms) for the detector with perfect resolution and the detector with 50 µm resolution is significant. At the bottom of Fig. 3, we show corresponding plots for a 100 mm gap between the GEM detector layers. Here the difference between perfect and 50 µm resolution is less dramatic as the larger gap between the GEMs smoothens the effect of the resolution on the muon track fitting.

## C. Reconstruction of the targets in the MT volume

For the reconstruction, we divide the MT volume into $10 \times 10 \times 5$ cm³ voxels. For each muon, we use a basic Point of Closest Approach (POCA) algorithm [8], [9] to get the 3D coordinates of a single scattering point ("POCA point") that is used to approximately represent what is actually the multiple scattering of the muon in the material. Every single voxel of the MT volume is subsequently assigned the mean value of the scattering angle of all the POCA points reconstructed within it [9]. In this way, we obtain a 3D tomographic image of the target materials within the MT volume based on the mean scattering angle. In Fig. 4 we show a horizontal slice (at z = 0 mm) of the reconstructed MT scenario described in Fig. 1 for different resolutions.

We evaluate how the detector resolution affects how accurately an image is reconstructed. We define a POCA accuracy ratio as the ratio between the number of POCA points reconstructed within a given target material in the MT station and the total number of muons scattered by that target material. This ratio is plotted in Fig. 5 for different resolutions and materials. The overall effect of a worse resolution on the reconstruction is to spread the POCA points for small scattering angle muons across the MT volume instead of localizing the POCA points where the interaction actually took place. This reduces the available statistics in the voxel associated with the material and deteriorates the imaging capability.

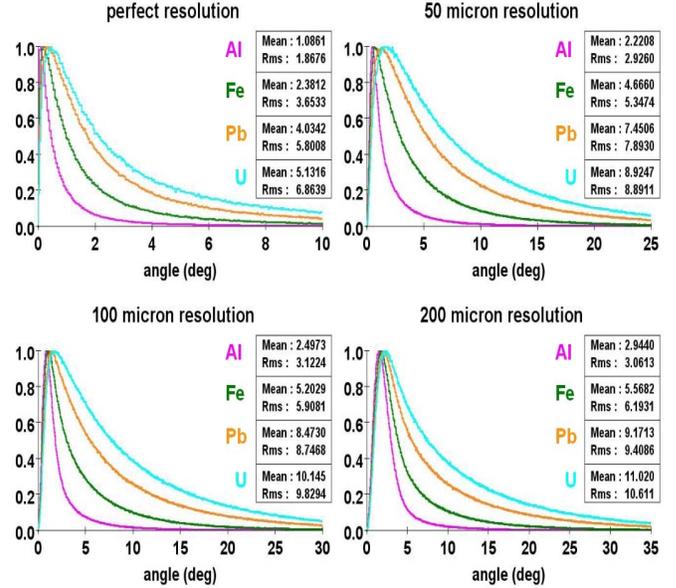

Fig. 2: Distribution of the scattering angles for cosmic ray muons traversing materials with different Z values in the MT volume for different detector resolutions.

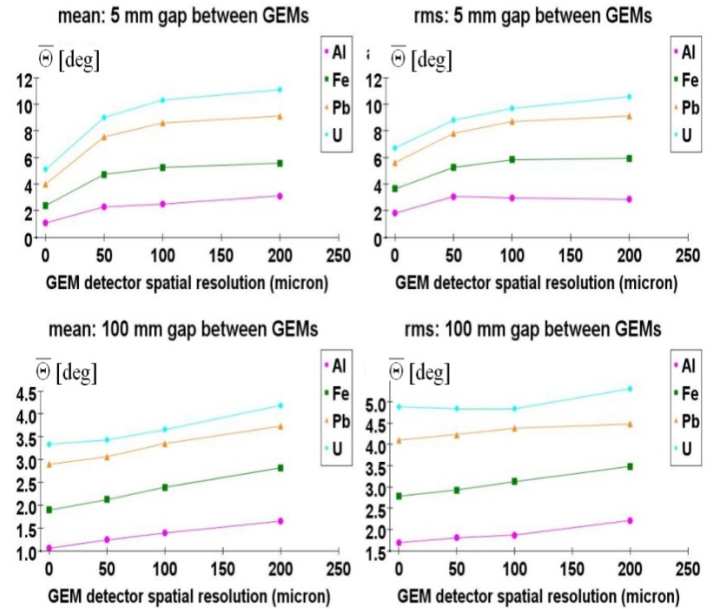

Fig. 3: Detector resolution effect on the measured mean and rms of the scattering angle distribution for 5 mm gap between the detectors (top) and 100 mm gap (bottom).

## III. Z-IDENTIFICATION TEST AND SIGNIFICANCE ANALYSIS

We study the ability to distinguish between low-Z, medium-Z, and high-Z materials as a function of exposure time and detector resolutions, which are the main parameters of the MT system for identifying high-Z threat materials in a cargo container mainly filled with low-Z and medium-Z materials. This study is done following the three steps described below.

### A. Selection criteria

We define the statistical criteria for identifying high-Z material in a voxel in the MT volume based on the comparison of the mean scattering angle $\overline{\Theta}_{voxel}$ of the material in that voxel and the mean scattering angle of a reference material $\overline{\Theta}_{ref}$. We first compute the mean scattering angle $\overline{\Theta}_{ref}$ for the material to be used as reference material for the analysis from our high statistics samples. Then we select the desired fixed confidence level (C. L.) for our selection test; here we select 99% C.L.

For every voxel in the MT volume, we define the interval of confidence around the mean scattering angle $\overline{\Theta}_{voxel}$ for the voxel as:

$$-\Theta_{voxel}^{cut} < \overline{\Theta}_{voxel} < \Theta_{voxel}^{cut} \quad (1), \text{ with}$$

$$\pm \Theta_{voxel}^{cut} = \overline{\Theta}_{voxel} \pm n_{voxel} \times \sigma_{\overline{\Theta}_{voxel}} \quad (2),$$

where $n_{voxel}$ is calculated using the Student's t distribution at 99% confidence level given the number of POCA points in the voxel and $\sigma_{\overline{\Theta}_{voxel}}$ is the uncertainty on the mean scattering angle $\overline{\Theta}_{voxel}$ in the voxel. The Student's t distribution is necessary because if the number of available POCA points is small, a simple Gaussian statistics would underestimate the width of the confidence interval.

We then define the significance of the selection test for each voxel and a given reference material as:

$$Sig_{voxel} = \frac{(\overline{\Theta}_{voxel} - \overline{\Theta}_{ref})}{\sigma_{\overline{\Theta}_{voxel}}} \quad (3)$$

A voxel is considered to be occupied by a reference material with 99% confidence if $|Sig_{voxel}| < n_{voxel}$.

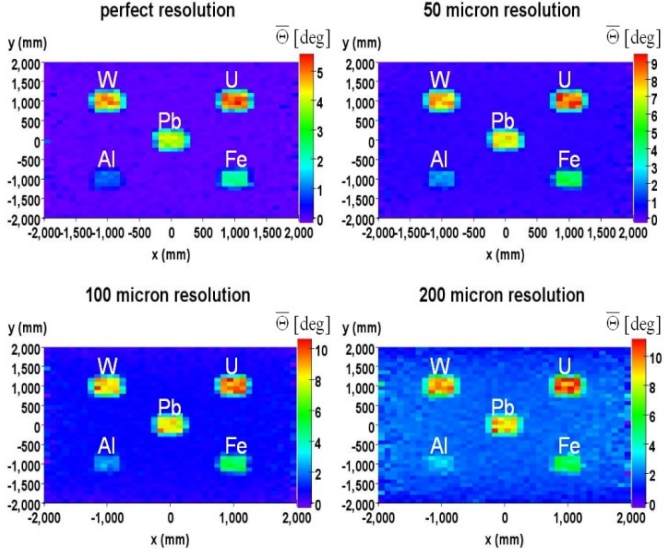

Fig. 4: Reconstruction of the 5-target scenario for different detector resolutions. The reconstruction is based on the mean angle of all POCA points in each voxel. The dimensions of the blocks are $40 \times 40 \times 10$ cm$^3$, the dimensions of the voxels are $10 \times 10 \times 5$ cm$^3$.

### B. Pre-selection of voxels with excess scattering above a given background

The first step is a pre-selection of those voxels inside the MT volume that are found to exhibit an excess of scattering over what is expected from typical medium-Z material. As our reference medium-Z material we choose the element Fe because Fe is expected to be a common medium-Z background in typical cargo, e.g. in the form of steel. All voxels containing materials with a Z value lower or equal to iron are considered background and the reference angle used for calculating $Sig_{voxel}$ is $\overline{\Theta}_{Fe}$. The pre-selection criterion is then $Sig_{voxel} > n_{voxel}$.

### C. Identification of Uranium in the MT volume

Next we test the hypothesis that the pre-selected voxels actually contain a particular threat material, e.g uranium. The reference now is U $(\overline{\Theta}_{ref} = \overline{\Theta}_{threat} = \overline{\Theta}_U)$ and the selection criteria for identifying U is now $|Sig_{voxel}| < n_{voxel}$.

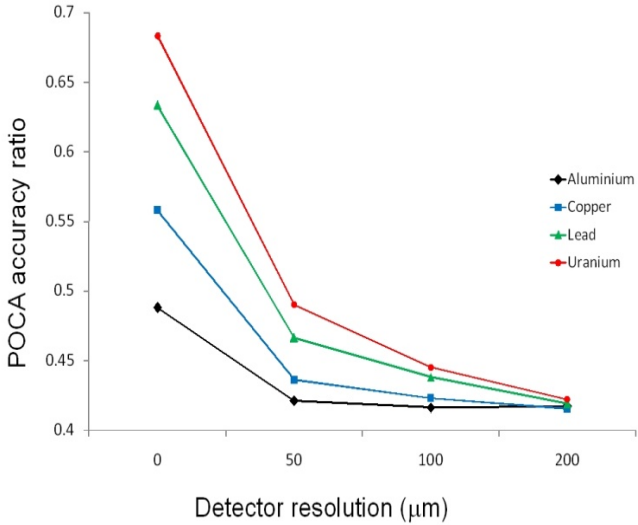

Fig. 5: POCA accuracy ratio for different materials. This is the ratio between the number of POCA points reconstructed in a volume occupied by a given target and the number of muons actually scattered by the target material.

### D. Uranium identification for large targets and 10 min measurement

Fig. 6 displays the significance $Sig_{voxel}$ of the $10 \times 10 \times 5$ cm$^3$ voxels that pass the high-Z pre-selection test after 10 min exposure and for different detector resolutions using the scenario from Fig.1. We clearly see significances higher than 5 at 99% confidence level for Pb, W, and U with 50 µm resolution. For worse resolutions, we still can pre-select all the voxels with a higher Z than Fe, although the significance decreases because of the loss of statistics (see Fig. 4).

Fig. 7 shows the results for the U hypothesis at 99% confidence level after 10 min exposure for different resolutions. In all cases, almost all U voxels are identified accurately. All voxels containing lead (Pb) are correctly rejected except when the resolution is 100 µm or worse. However, we observe false positive selection of some of the voxels containing tungsten even for a detector with perfect resolution (green circles).

### E. Uranium identification for large targets and 1 min measurement

Fig. 8 displays the significance $Sig_{voxel}$ of the voxels that pass the high-Z pre-selection test if the exposure time is reduced to 1 min, i.e. for a statistics 10 times lower. For a detector with perfect resolution, the targets with Z higher than Fe are only partially identified and many voxels are missing. For 50 µm resolution detectors, we are still able to partially identify the U block, but only one or two voxels are selected for W and Pb. For worse resolution we can barely discriminate any materials (even U) against Fe background.

The U hypothesis test for 1 min. exposure time reveals a partially reconstructed U block for perfect resolution and 50 µm resolution detectors (Fig. 9), but also W and Pb voxels reconstructed as false positive signals (green circles).

### F. Uranium identification for small targets and 10 min measurement

We repeat the simulation and discrimination test but this time with small liter-sized targets ($10 \times 10 \times 10$ cm$^3$) and $5 \times 5 \times 5$ cm$^3$ voxels. The significance analysis (Fig. 10) shows that we can clearly identify the excess above Fe background after 10 min measurement for a detector resolution up to 100 µm. For 200 µm, no voxel is identified as containing material with a Z value higher than Fe. The U hypothesis (see Fig. 11) shows all the W and Pb voxels as false positive signals (green circles). This means that we cannot discriminate between Pb, W, and U in this case. For 200 µm resolution only one voxel is identified as corresponding to U, which constitutes a false negative signal (red circle in Fig. 11), i.e. in this case the threat material would go undetected. These results are mainly due to the low statistics when the target and voxel size is small.

## IV. SYSTEMATIC EFFECTS

The 3D tomographic image of the MT volume is based on the mean value $\overline{\theta}_{voxel}$ of the measured scattering angles at all the scattering points given by POCA in each voxel of the MT volume. The calculated $\overline{\theta}_{voxel}$ for a material of a given Z value depend on many parameters, e.g. the dimensions of the target, the geometry of the MT volume, the location of the target in the MT volume, and the spatial resolution of the detector. For the discrimination test reported in section III, we need to have reference data samples that take into account these systematic effects for each material. We report a study of these different systematic effects on the MT performance in this section.

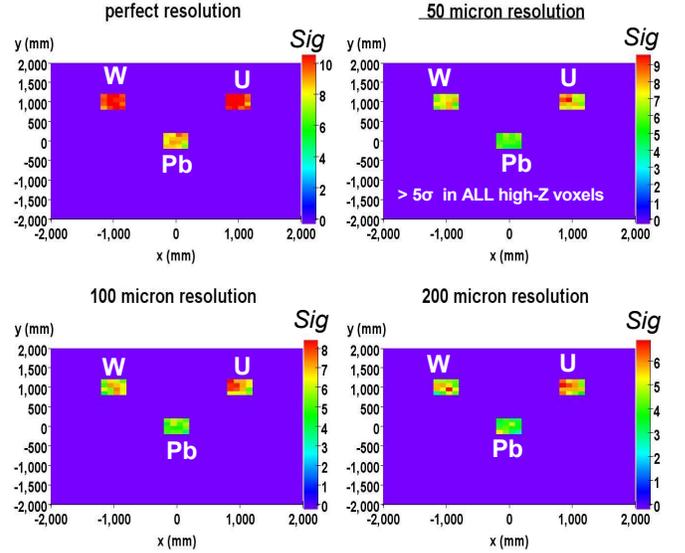

Fig. 6: Significance of the excess over medium-Z background material (Fe) based on the mean scattering angle at 99% confidence level after 10 min measurement for the scenario from Fig.1 and for different detector resolutions.

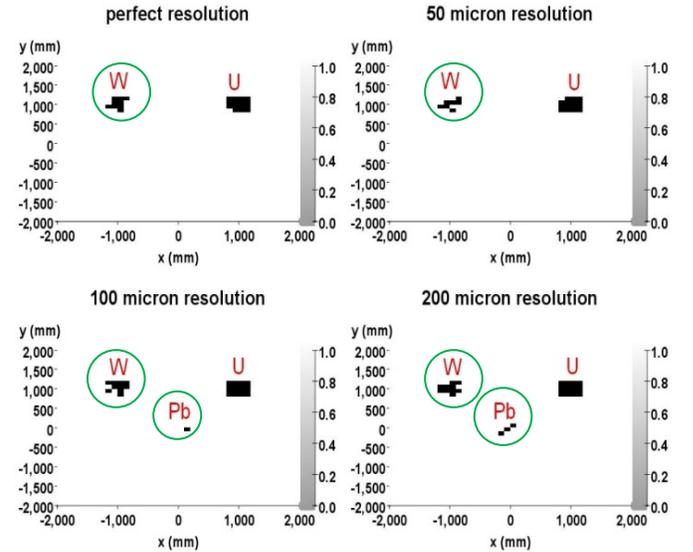

Fig. 7: Uranium hypothesis test: Identification of pre-selected voxels as uranium at 99% confidence level after 10 min measurement and for different detector resolutions. Green circles indicate voxels with false positive identification.

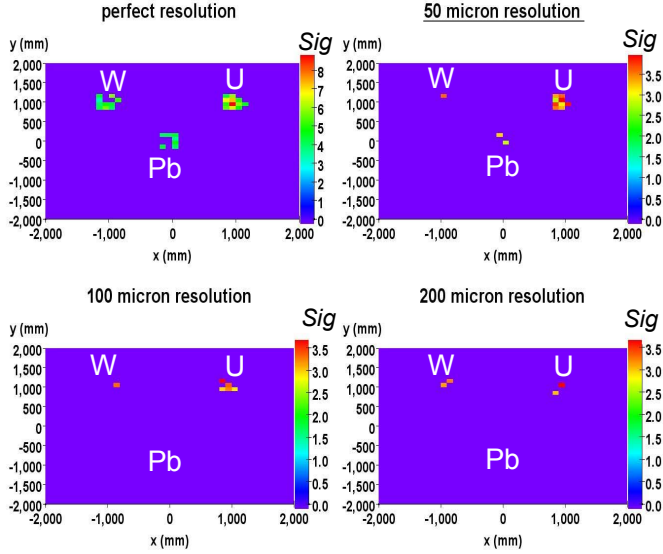

Fig. 8: Same as Fig. 6, but for 1 min measurement.

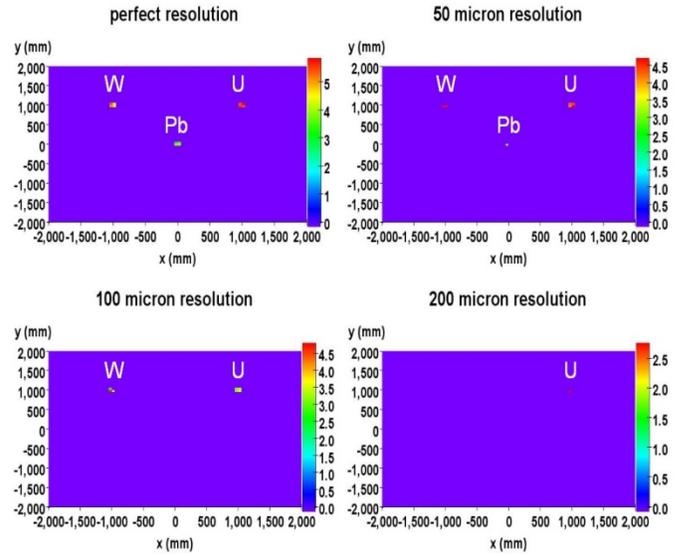

Fig. 10: Same as Fig. 6, but for $10 \times 10 \times 10$ cm$^3$ material blocks.

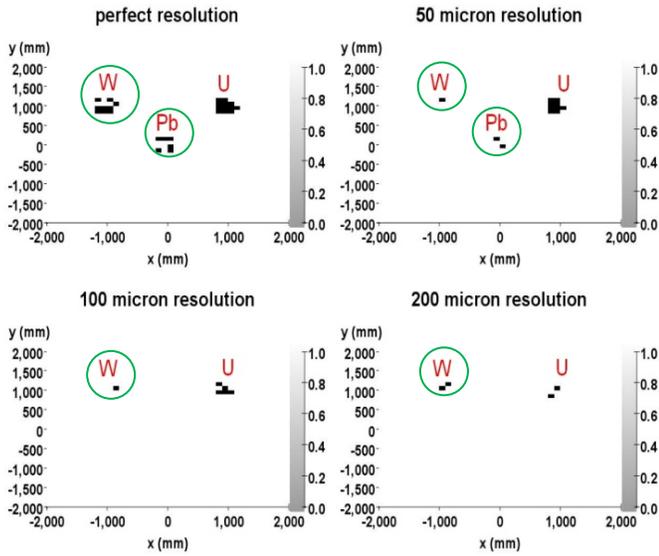

Fig. 9: Uranium hypothesis test: Same as Fig. 7, but for 1 min measurement. Green circles indicate voxels with false positive identification.

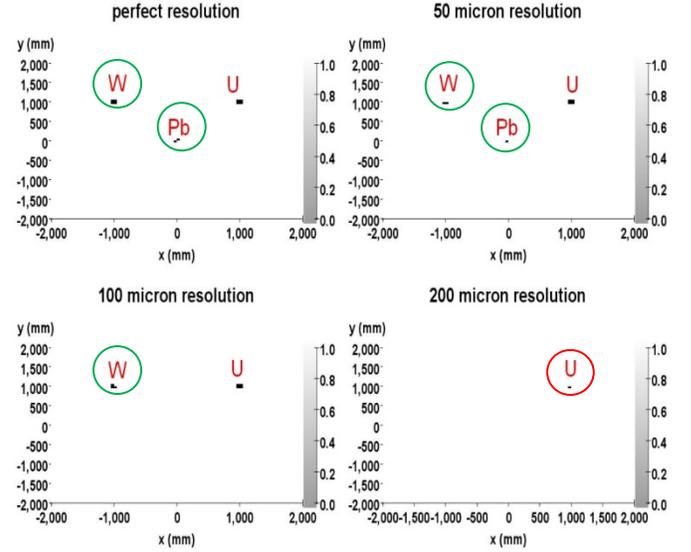

Fig. 11: Uranium hypothesis test: Same as Fig. 7, but with $10 \times 10 \times 10$ cm$^3$ targets. Green circles indicate voxels with false positive identification. The red circle indicates a case of false negative identification.

### A. Effect of the target dimensions

The first effect is due to the dimensions of the target we are trying to identify. The mean scattering angle $\overline{\Theta}_{mat}$ for a Z material depends on the radiation length $\lambda_{mat}$ of this material and consequently on the thickness and the size of the target since the cosmic ray muons come from different angles at different energies. Fig. 12 shows for different materials how the mean scattering angle $\overline{\Theta}_{mat}$ (right) as well as the rms of the distribution (left) increases with the size (top) and thickness (bottom) of the target.

### B. Effect of the target position within the MT volume

Because of the geometry of the MT and the selection of only those muons that can be detected by a pair of tracking stations, the mean scattering angle $\overline{\Theta}_{mat}$ for a given material varies slightly with the position of the material within the MT volume. As shown in Fig. 13, a difference can be seen in the reconstructed mean scattering angles for a target placed in the center relative to a target at the edge of the horizontal plane. The study is repeated with targets in the horizontal plane at different z along the vertical axis in the MT.

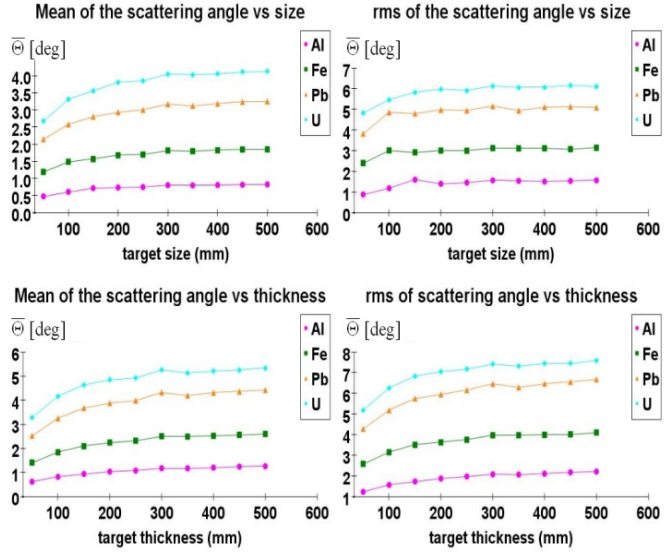

Fig. 12: The effect of the target dimensions on the scattering angle distributions for different materials. The two top plots show the mean angle (left) and the rms (right) of the distribution vs. the x-y size of the targets. The bottom plots show the mean (left) and rms (right) vs. the target thickness in z.

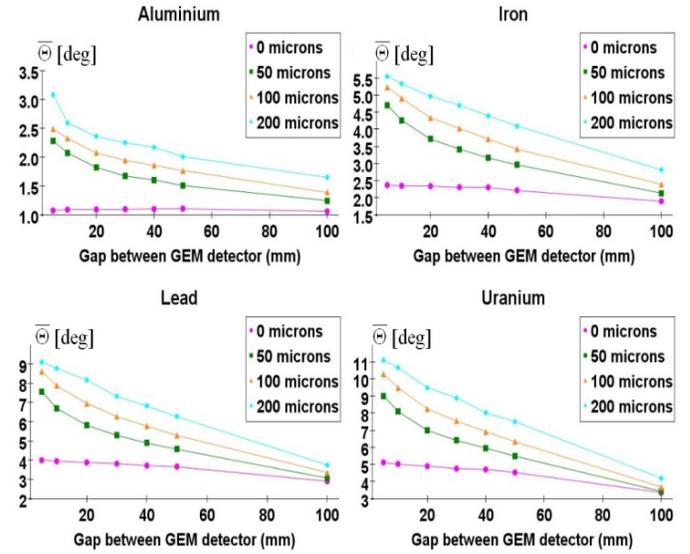

Fig. 14: Effect of the gap between the GEM detectors. For 100 mm gap and for high-Z material, the resolution effect on the measured mean angle is suppressed for resolutions less than 100 μm, making the MT performances less sensitive to the detector resolution.

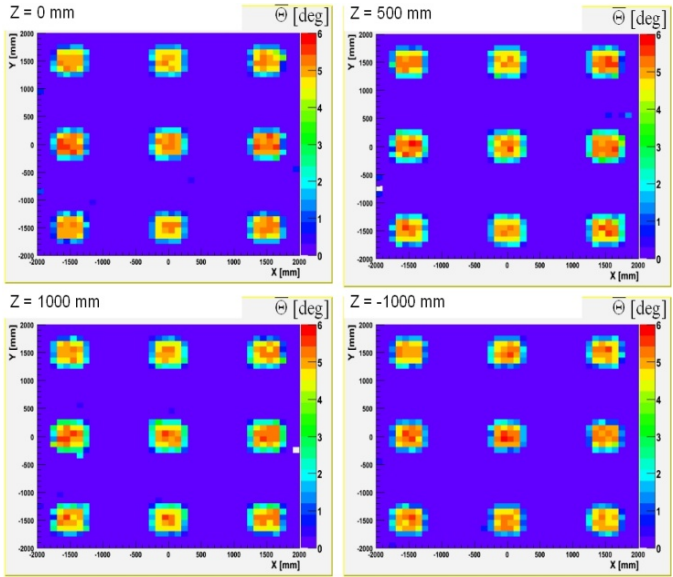

Fig. 13: Reconstructed mean angle for 4 simulated scenarios. Each scenario has 9 identical U targets at different locations in the X-Y plane and at a given position along the vertical axis (z = -1000 mm, 0, 500 mm, 1000 mm) in the MT volume. We can clearly see an effect of the position of the target on the mean angle.

### C. Effect of gap between detectors

The mean value of the scattering angle $\overline{\Theta}_{mat}$ calculated from the tracks of the incoming and outgoing muons increases with the increasing detector resolution. An additional effect of the resolution concerns the accuracy of the POCA localization of the interaction point. The drop of the POCA accuracy ratio discussed previously (see Fig. 4) is mainly due to muons that are scattered with small scattering angle and for which the scattering point is incorrectly reconstructed far outside the actual interaction region in the material.

Consequently, the mean scattering angle and the rms of the voxels representing these materials are strongly shifted towards higher value.

If the gap between detector layers in the tracking stations is increased, the influence of the detector resolution on the muon track is reduced due to the increase in lever arm for the track. Fig. 14 shows that increasing the detector gap consequently lowers the mean angle so that the mean angle for finite resolutions converges towards the mean angle of a detector with perfect resolution. This gives a possible handle on improving detector performance if the desired 50 μm resolution cannot be achieved experimentally, albeit at the cost of making the tracking stations less compact.

### V. CONCLUSION

This study clearly shows that cosmic ray muon tomography can discriminate sensitive high-Z nuclear material such as uranium against iron or steel background with high statistical significance if the detectors have good spatial resolution of ~100 μm and if the targets are exposed for 10 minutes. In this case, uranium can even be discriminated against lead, a high-Z material, with 99% confidence. For detectors with even better spatial resolution of ~50 μm and with optimized tracking station geometry, we expect to reach similar performance for a much shorter exposure time on the order of 1 minute. Various systematic effects are observed, but do not detract from this overall conclusion. Consequently, we plan to construct a small prototype of a muon tomography station with GEM detectors to confirm this expectation experimentally.

ACKNOWLEDGMENT AND DISCLAIMER

This material is based upon work supported in part by the U.S. Department of Homeland Security under Grant Award Number 2007-DN-077-ER0006-02. The views and conclusions contained in this document are those of the authors and should not be interpreted as necessarily representing the official policies, either expressed or implied, of the U.S. Department of Homeland Security.